# Central Diffraction at the LHCb

Jerry W. Lämsä[1,2] and Risto Orava[1]

[1]Helsinki Insitute of Physics, and Division of Elementary Particle Physics, Department of Physics,
PL 64 (Gustaf Hällströmin katu 2a), FI-00014 University of Helsinki, Finland
[2]Physics Department, Iowa State University, Ames, 50011 Iowa, U.S.A.

**Abstract**

The LHCb experiment is shown to be ideal for studies of exclusive final states from central diffractive reactions. The gluon-rich environment of the central system allows detailed QCD studies and searches for exotic meson states, such as glueballs, molecules, hybrids and new charmonium-like states. It would also provide a good testing ground for detailed studies of heavy quarkonia. Due to its distinct design features, the LHCb can accurately measure the low-mass central systems with good purity. The efficiency of the Forward Shower Counter (FSC) system for detecting rapidity gaps is shown to be adequate for the proposed studies. With this detector arrangement, valuable new data can be obtained by tagging central diffractive processes.

# 1 Introduction

A unique central diffraction physics program can be realized with the LHCb experiment[1]. Low-mass central diffractive final states decaying into a small number of particles, with no forward secondary (shower) particles produced from interactions in the beam pipes, can be selected by adding Forward Shower Counters (FSCs) [1] on both sides of the LHCb experiment at the LHC Interaction Point (IP8).

The central diffractive (CD) reaction

$$p\,p \rightarrow p + M + p, \qquad (1)$$

where M is a hadronic state, is selected with the use of FSCs, which define the rapidity-gaps (denoted +).[2] Detection of the outgoing protons, which will mostly remain within the beam pipe, is not required for this study. The reaction is considered to be mediated by the two-gluon colour-singlet interaction

$$g\,g \rightarrow M. \qquad (2)$$

The physics program will be primarily concerned with a search for the production of meson states such as glueballs, hybrids, and heavy quarkonia $\chi_c$, $\chi_b$ [2,3]. A search will be made for structure in the mass spectra of exclusive decay states of M, such as, $\pi^+\pi^-$, $K^+K^-$, $2\pi^+2\pi^-$ and $K^+K^-\pi^+\pi^-$, $K^+K^o\pi^-$, $K^-K^o\pi^+$, $p\bar{p}$, $\Lambda\bar{\Lambda}$ and others. A strong coupling for the reaction $g\,g \rightarrow M$ is expected as a result of the two-gluon exchange. Being central to QCD, discovery of glueballs would be of great importance. The very high statistics studies of the process in Eq. 2 provides this possibility.

For the case of pomeron – pomeron interactions, the central system is dominantly produced with spin-parity $J^{PC} = 0^{++}, 2^{++}$, etc. The decays with low multiplicities such as $\pi^+\pi^-$ or $K^+K^-$ can be used as efficient spin-parity analyzers [2]. The t-channel exchanges over the large rapidity gaps can only be colour singlets with Q = 0. Known exchanges are the photon $\gamma$ and the pomeron P.[3] Another possible, but not yet observed, exchange in QCD is the odderon, O, a negative C-parity partner to the P with at least 3 gluons. The physics programme includes sensitivity to odderon exchange. Double pomeron exchange (DPE) produces primarily $I^G J^{PC} = 0^+ 0^{++}$ states, with some $0^+ 2^{++}$ admixture. $J^{PC} = 1^{--}$ states such as J/$\psi$ and Y are produced by $\gamma$P, but can also be produced by OP.

The gluon-rich and quantum number filtered central system is a laboratory for studies of QCD and glueball spectroscopy. Studies on production rates of $\eta$, $\eta'$ mesons, baryons, etc. could be compared with inclusive p$\bar{\text{p}}$ interactions at $\sqrt{s} \approx M$, where M is the mass of the central system [2].

---

[1] The early LHC program, with anticipated 5 TeV + 5 TeV proton – proton interactions, at a low luminosity (no pile-up), is considered in this study.
[2] Concerning the purity of the exlusive central diffractive process, an earlier study by the authors is referenced, see [1].
[3] The gluon passes the Q requirement, but is not a colour singlet; however one or more additional gluons can cancel its colour and form a Pomeron.

Based on forward shower counters, the CDF experiment at the Tevatron has recently produced a series of highly valuable results on central exclusive production of $\chi_c$, $\gamma\gamma$, di-jets and J/$\psi$ [4].

## 2 Experimental Overview

For the rapidity gap detection, FSC scintillation counters would be employed surrounding the beam pipes in the region from 20m to 100m on both sides of the LHC interaction point (IP8), see Fig.1. The FSCs will detect showers from very forward particles interacting in the beam pipe and surrounding material. The absence of a shower indicates a rapidity gap.

The LHCb RICH detectors provide particle identification over a wide momentum range. A mass resolution of $\sim$ 20 MeV is expected for low multiplicity states [5].

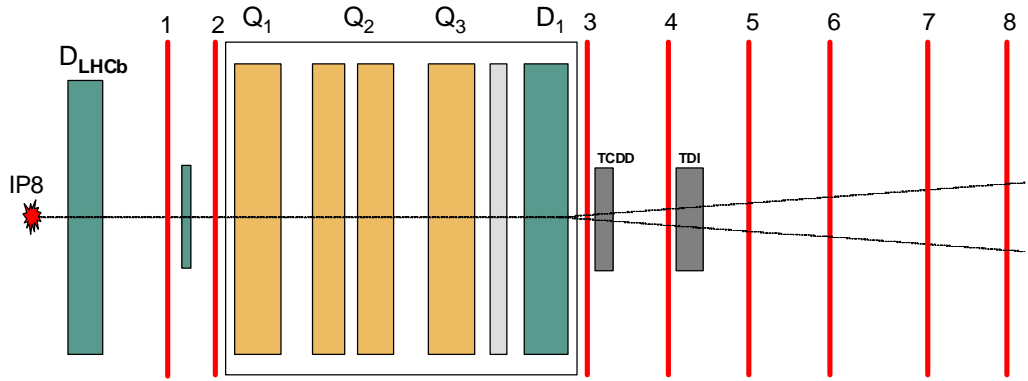

**Figure 1.** *The layout of LHCb detectors at the LHC Interaction Point (IP8). The proposed Forward Shower Counters (FSCs) are shown as vertical lines (1 to 8). The locations of the dipole (D) and quadrupole (Q) magnet elements are shown as green (dark) and yellow (light) boxes.*

To detect a low-multiplicity decay, a small number (e.g., less than 5) of charged tracks are required to strike the Scintillator Pad Detector (SPD) [5]. Finally, the LHCb VErtex LOcator (VELO) would be required to have an absence of charged tracks within a given polar angle range, i.e., a central angle veto. For the present study, the angular range chosen was 10 to 170 deg.[4] The results are not sensitive to the exact range used. Monte Carlo simulation of the detector, beam pipe and magnet elements has been done with GEANT [6]. It is also to be noted that an average of one interaction per beam crossing is expected with the standard LHCb low luminosity running.

---

[4] Alternatively, shower counters surrounding the VELO region could also be considered.

In summary, the requirement for a trigger would be a low charged multiplicity in the SPD, a restriction on charged tracks in the VELO, and an absence of a signal in the FSCs.[5]

## 3 Central Diffraction (CD) Acceptance

The CD reaction, Eq.(1), was simulated with PHOJET 1.1 [7]. The decay of the central system M into low-multiplicity exclusive final states was generated isotropically with PYTHIA 6.2 [8]. The states $\pi^+\pi^-$, $K^+K^-$, $2\pi^+2\pi^-$ and $K^+K^-\pi^+\pi^-$, are considered for this paper. The spectrometer detector angular acceptance region was taken as within < 250 mr (vertical), < 300 mr (horizontal), and > 25 mr. The detection efficiencies for events (in the forward direction) were calculated as a function of the diffractive mass, and are shown in Fig.2. For the final states considered, the acceptances range between 10 and 20 percent, for M < 4 GeV. For higher masses, the acceptances increase gradually.

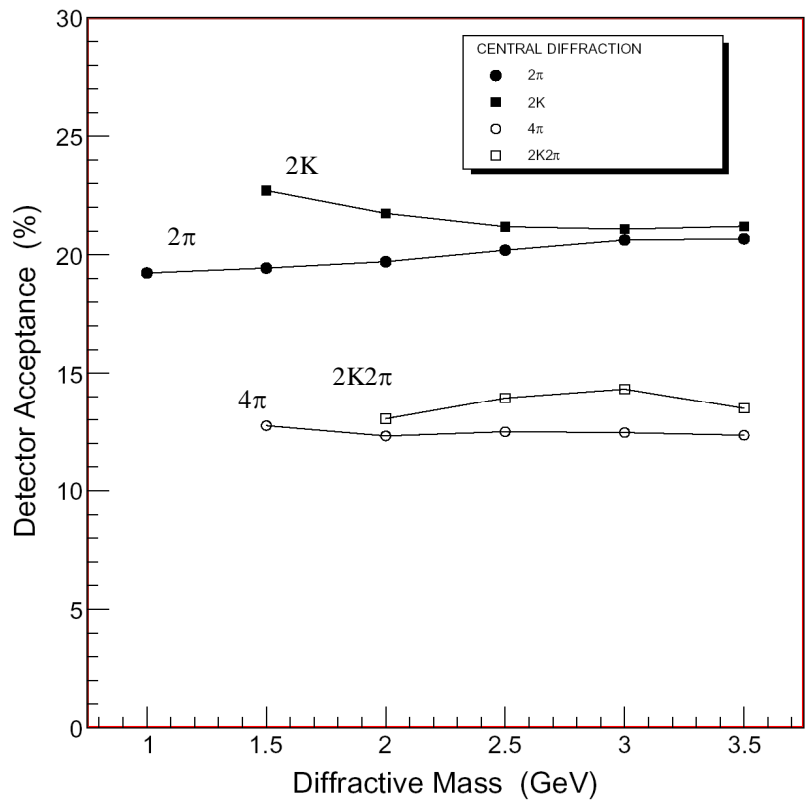

**Figure 2.** *The detector acceptance as a function of the central diffractive mass for $\pi^+\pi^-$, $K^+K^-$, $2\pi^+2\pi^-$ and $K^+K^-\pi^+\pi^-$ decay channels.*

---

[5] Additional conditions could also involve the detection of neutral particles.

# 4 Single Diffraction (SD) Background

Background from single diffraction (SD) is a concern since the multiplicities are mainly forward and will often have a small number of charged particles (less than 5) in the detector acceptance satisfying the SPD trigger requirement. The SD events were generated with PYTHIA. The probability per event that a given number of charged particles fall within the detector acceptance region, as defined above, is shown as the top curve (with filled circles) in Fig.3.

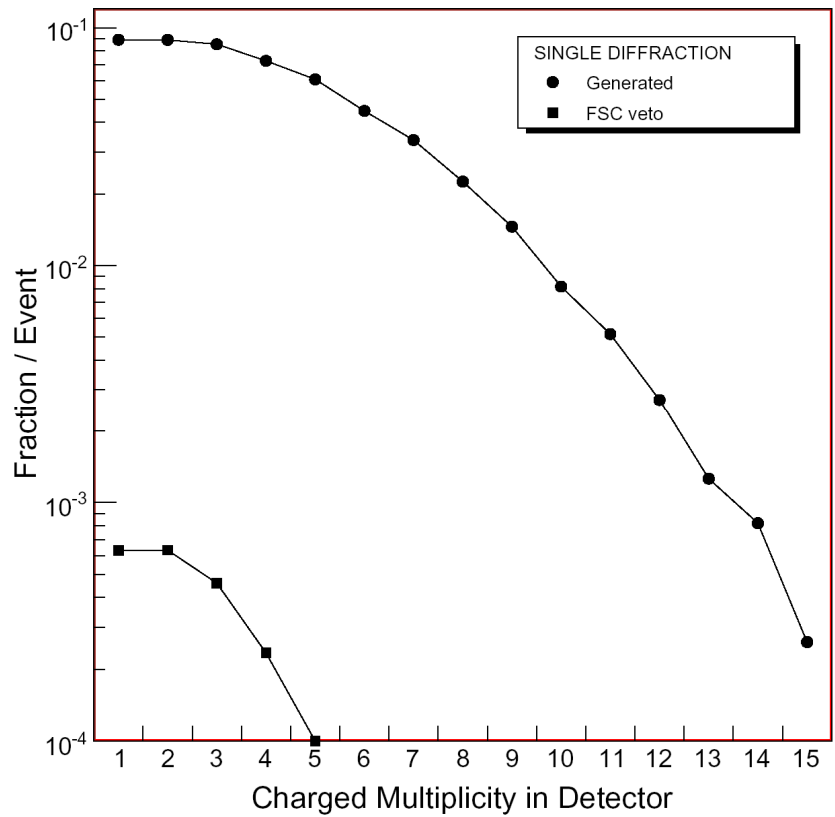

**Figure 3.** *The probability per event for a given number of charged particles to be emitted within the spectrometer detector acceptance region is given by the upper curve (filled circles), the lower curve (filled squares) gives the acceptance with deployment of the FSCs.*

The value for multiplicities less than five is about 9%. This represents a rather large background which needs to be suppressed. An example of the contribution from this background to the $\pi^+\pi^-$ and $K^+K^-$ mass distributions where a unique $\pi^+\pi^-$ and $K^+K^-$ pair is within the detector acceptance, is shown in Figs.4 and 5.

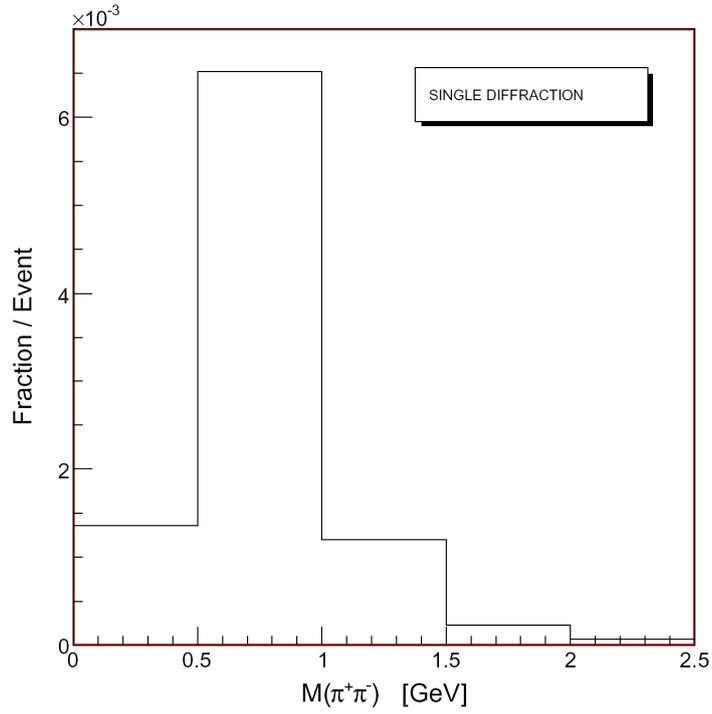

**Figure 4**. *The contribution from single diffractive events that produce a unique $\pi^+\pi^-$ pair within the detector acceptance, as a function of the $\pi^+\pi^-$ effective mass.*

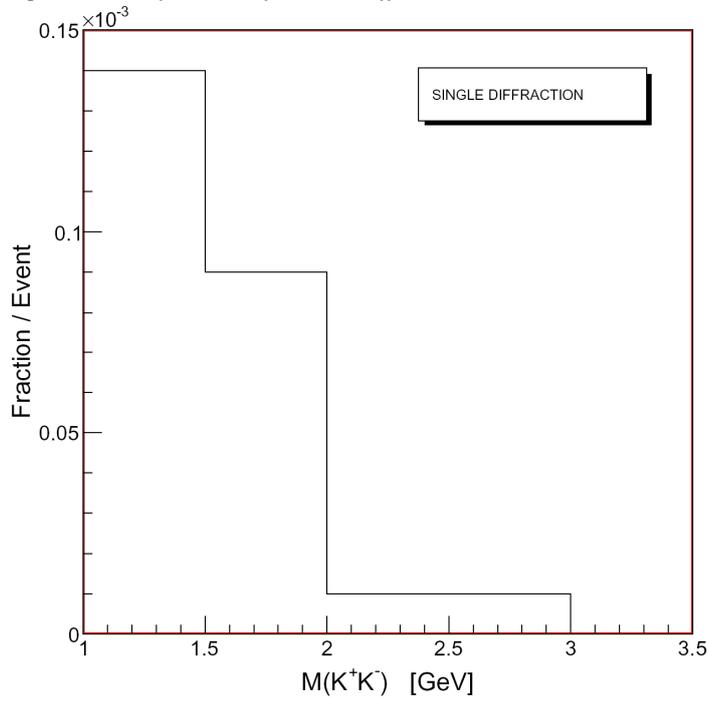

**Figure 5.** *The contribution from single diffractive events that produce a unique $K^+K^-$ pair within the detector acceptance, as a function of the $K^+K^-$ effective mass.*

The deployment of FSCs will allow background to be reduced. Single diffraction events will generally produce shower particles from interactions in the beam pipe.[6] The FSC detection efficiencies for single diffraction events were calculated as a function of the diffractive mass, see Fig.6. At least five hits in any of the FSCs has been required. As described earlier, the central angle veto was included. The efficiency is high at the larger masses. Particles from the smallest masses with their low multiplicities rarely enter the LHCb detector.

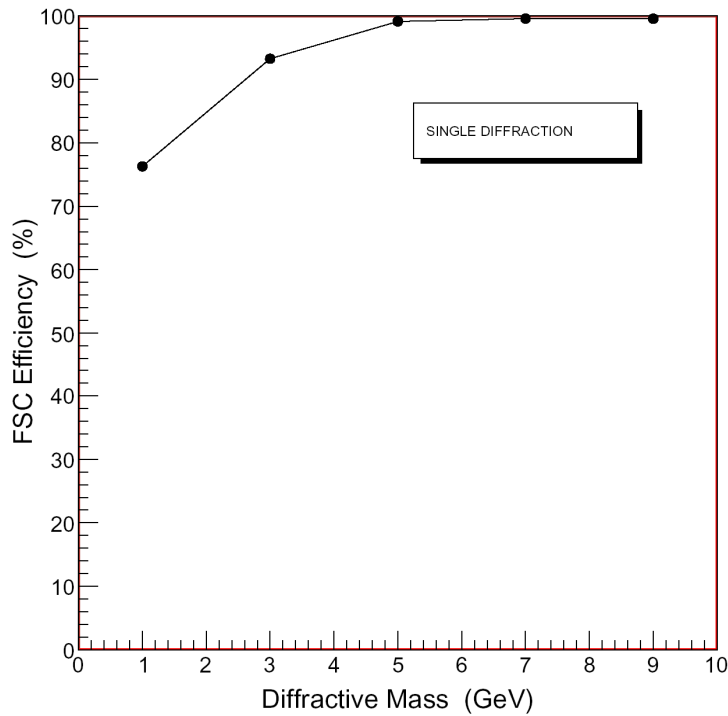

**Figure 6.** *The efficiency to detect single diffractive events (SD) by the Forward Shower Counters (FSCs) as a function of the diffractive mass.*

With the above FSC requirement, the probability per event, that a given number of charged particles enter the detector acceptance region is shown as the bottom curve with filled squares in Fig. 3. The background due to the SD events is reduced by the FSCs by more than two orders of magnitude.[7]

## 5 Non-Diffraction (ND) Background

The analysis is similar to the SD study with the non-diffractive (ND) events generated by PYTHIA. The probability per event, that a given number of charged particles fall within the detector acceptance region, as defined above, is shown in Fig.7. The probability for a small number (less than five) of charged particles is small. This background can be reduced with the FSCs.

---

[6] The low-mass single diffractive events provide an interesting luminosity monitor, see Ref. [9].
[7] For further discussion, see Ref. [1].

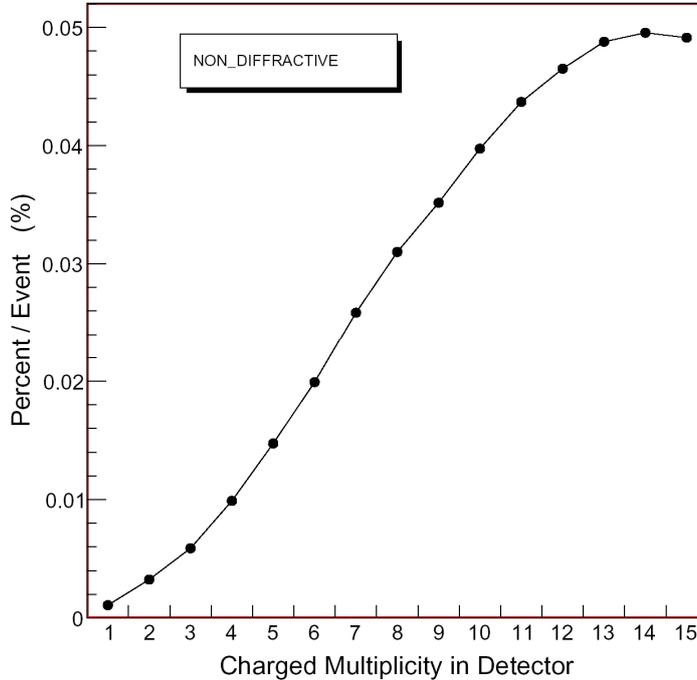

**Figure 7.** *The efficiency to detect non-diffractive events (ND) by the Forward Shower Counters (FSCs) as a function of the charged multiplicity in the detector.*

As in the SD case, the ND events invariably produce shower particles from interactions in the beam pipes. The FSC detection efficiencies for the non-diffractive events were calculated as a function of the charged multiplicity in the detector. At least five hits in any of the FSCs has been required. In addition, the central angle veto was included. The efficiency was essentially 100% over the multiplicity spectrum for the $10^6$ events generated (i.e., all $10^6$ events were detected).

## 6 Central Diffraction Purity

The most important background to the study of the CD exclusive low-multiplicity final states is expected to come from the production and decay of higher mass and multiplicity CD states. The background to the exclusive states resulting from 'feed-down' of higher-mass final-states was calculated with PHOJET. The fraction of cases (purity), where a particle combination, $\pi^+\pi^-$, $K^+K^-$, $2\pi^+2\pi^-$, (within the acceptance) originates from the exclusive decay of the central system M *rather than* from feed-down of higher mass states, is presented in Fig.8.

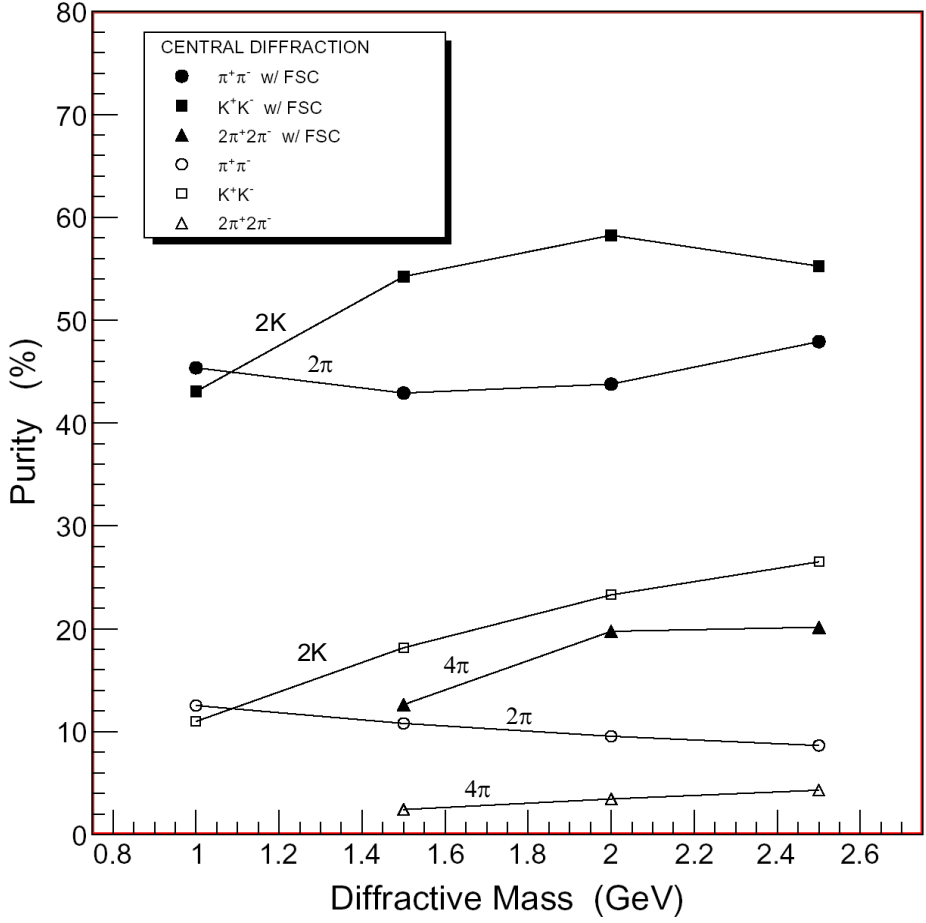

**Figure 8.** *The purities, i.e. the fractions of cases where a particle combination, $\pi^+\pi^-$, $K^+K^-$, $2\pi^+2\pi^-$, (within the acceptance) originates from the exclusive decay of the central system M rather than from feed-down of higher mass states, as a function of the effective mass of the particle combination..*

The purities are below 30%. The influence of the FSC detection efficiencies and central angle veto, for the CD events (by PHOJET) were calculated as a function of the diffractive mass. When the requirement of the presence of rapidity gaps is added, i.e., absence of shower particles in the FSCs, the purities increase to ~50% for $\pi^+\pi^-$ and $K^+K^-$ and to ~20% for $2\pi^+2\pi^-$. These purities are acceptable for the proposed studies.

# 7 Conclusions

Feasibility studies of the exclusive central diffractive processes for the LHCb experiment have been carried out.[8] With a simple addition of Forward Shower Counters (FSCs), the experiment is shown to be ideally suited for detailed QCD studies and searches for exotic meson states, such as glueballs, molecules, hybrids and heavy quarkonia.

## Acknowledgements

Invaluable advice from Werner Herr and Valery Khoze, and financial support by the Academy of Finland are gratefully acknowledged.

---

[8] The authors have carried out an earlier feasibility study of forward physics potential of the ATLAS and CMS experiments, see Ref. [10, 1].